\documentclass[twoside,twocolumn,english,superscriptaddres,showpacs,longbibliography,aps,prb]{revtex4-2}

\usepackage[T1]{fontenc}
\usepackage[utf8]{inputenc}
\usepackage[english]{babel}
\usepackage{bm}
\usepackage{amsmath}
\usepackage{amssymb}
\usepackage{graphicx}
\usepackage[unicode=true, pdfusetitle, bookmarks=true,bookmarksnumbered=true,bookmarksopen=true,bookmarksopenlevel=1,
breaklinks=false,pdfborder={0 0 0},pdfborderstyle={},backref=false,color links, citecolor=blue,linkcolor=red,urlcolor=blue] {hyperref}
\usepackage[nameinlink]{cleveref}
\Crefname{figure}{Fig.}{}

\makeatletter
\def\set@firstnote#1{%
 \@ifnum{\firstnote@num=#1\relax}{}{%
  \class@warn@end{Endnote numbers changed: rerun LaTeX}%
 }%
 \immediate\write\@mainaux{%
   \global\mathchardef\string\firstnote@num#1\relax
 }%
}%

\usepackage{braket}
\usepackage{bm}
\usepackage{tikz}
\usepackage{pgfplots}
\usepackage{pgf}
\usepackage{subfigure}
\usepackage{nicematrix}
\usepackage{diagbox}
\usepackage{orcidlink}

\renewcommand{\selectlanguage}[1]{}

\makeatother

\pgfplotsset{compat=1.18}
\begin{document}

\title{Operator-state correspondence in simple current extended conformal field theories: Toward a general understanding of chiral conformal field theories and topological orders}
\author{Yoshiki Fukusumi}
\author{Guangyue Ji}
\author{Bo Yang} 
\affiliation{Division of Physics and Applied Physics, Nanyang Technological University, Singapore 637371.}

\date{\today}
\begin{abstract}
In this work, we revisit the operator-state correspondence in the Majorana conformal field theory (CFT) with emphasis on its semion representation. Whereas the semion representation (or $Z_{2}$ extension of the chiral Ising CFT) gives a concise ``abelian" (or invertible) representation in the level of fusion rule and quantum states, there exists subtlety when considering the chiral multipoint correlation function. In this sense, the operator-state correspondence in the semion sector of the fermionic theory inevitably contains difficulty coming from its anomalous conformal dimension $1/16$ as a $Z_{2}$ symmetry operator. By analyzing the asymptotic behaviors of the existing correlation functions, we propose a nontrivial correspondence between the chiral conformal blocks and bulk correlation functions containing both order and disorder fields. One can generalize this understanding to $Z_{N}$ models or fractional supersymmetric models (in which there exist long-standing open problems). We expect this may improve our understanding of the simple current extension of CFT which can appear commonly in the studies of topologically ordered systems.
\end{abstract}

\maketitle

\section{Introduction}

Boson-fermion correspondence is one of the most celebrated frameworks for constructing and analyzing a wide class of lattice and field-theoretic models. One can see the profound history, initiated by Onsager's solution of the Ising model\cite{Onsager:1943jn}, bosonization of fermionic models\cite{Coleman:1974bu,Mandelstam:1975hb,Witten:1983ar}, quark models \cite{Goddard:1984hg,Goddard:1984vk,Goddard:1986ee,Goddard:1988md} and so on. We also note the significance of the fermionic representation and its close relation to integrability and renormalization group\cite{PhysRevB.13.316,Zamolodchikov:1989hfa}. By interpreting fermionization as $Z_{2}$ simple current extension, one can observe that $Z_{2}$ symmetry extension and its $Z_{2}$ anomaly classification give a concise RG understanding of the Haldane conjecture\cite{Furuya:2015coa,Fukusumi:2022xxe,Fukusumi_2022_c,Fukusumi_2022}. Based on this modern understanding, one can relate the integrability of a theory with the absence of its $Z_{2}$ anomaly and the corresponding fermionization.

In this work, we revisit the structure of the simplest fermionic model, the Majorana fermionic conformal field theory (CFT). For this model, the structure of the correlation functions has been studied many times, and one can extensively check its fermionized representation which has captured attention recently\cite{Runkel:2020zgg,Runkel:2022fzi,Hsieh:2020uwb,Seiberg:2023cdc}. It might be surprising but there still exists a lot to reconsider when analyzing the operator-state correspondence in this model. More specifically, we have found a nontrivial correspondence between the chiral conformal block defined by the solution of the Belavin-Polyakov-Zamolodchikov (BPZ) equation\cite{Belavin:1984vu} and their asymptotic behaviors, and the bulk correlation functions constructed from local bulk operators and the disorder operator. More precisely, this is a correspondence between the chiral CFT and the Schottky double of bulk CFT, and we call this correspondence the CCFT/DCFT correspondence. This can be thought of as an expression of the $\text{CFT}_{D}/\text{BQFT}_{D+1}$ correspondence (or $\text{CFT}_{D}/\text{BTQFT}_{D+1}$ correspondence), known as the bulk-edge correspondence\cite{Bernevig2008PropertiesON,Bernevig_2008,Fukusumi:2022xxe}. For example, in a fractional quantum Hall (FQH) system, our framework may enable one to generalize Laughlin's flux attachment argument even to the systems with non-abelian anyons. More specifically, we demonstrate the interpretation of conformal blocks and the corresponding wavefunctions based on both boson and fermion representations by identifying the set of fundamental operators. Whereas they are one of the most fundamental objects in chiral CFT and the corresponding TQFT, it might be surprising for the readers, that our method is the first time to write down them directly.

Our work gives a simple understanding of the conformal blocks appearing in modern condensed matter physics, statistical mechanical physics and mathematical physics, based on fermionization or on the language of ``particle physics" (i.e. the quark model in the older terminology). We expect similar nontrivial operator-state correspondence to appear in a general simple current extension of CFT \cite{Gato-Rivera:1991bqv,Fuchs:1996dd}  which can be thought of as the building block of topologically ordered systems\cite{Ino:1998by,Cappelli:2010jv,Lu_2010,Schoutens:2015uia,Fuji_2017,Fukusumi_2022_c,Henderson:2023kjz}(However, the rigourous formulation of the simple current or  $G$-symmetry extension of a chiral conformal field theory is a difficult problem also in mathematical studies. We note several recent works in this research direction\cite{etingof2009weaklygrouptheoreticalsolvablefusion,Barkeshli:2014cna,Babakhani_2023,Galindo:2024qzg,Gannon:2024tcl}. As we demonstrate in the manuscript, the breaking of the naive operator-state correspondence is a source of the difficulty.). In particular in the context of non-abelian Moore-Read states in the fractional quantum Hall systems, we can write down many-body states with four or more quasihole with non-abelian anyons with unambiguous operator-state correspondence. It should be noted that this interpretation of the wavefunctions is not obvious from the original interpretation with the Ising CFT or (bosonic) $M(3,4)$ minimal model. Moreover, the simple current extension of CFT can be thought of as a natural generalization of Majorana CFT and gives a lot of information on its renormalization group behaviors (For the modern aspect of this kind of studies, see \cite{Lecheminant:2015iga,Kikuchi:2022ipr}, for example.).

The rest of the manuscript is organized as follows. In Sec.\ref{fusion_section}, we revisit the structure of the Ising and Majorana CFT with emphasis on operator-state correspondence and fusion rules. A modern view of {boson-fermion correspondence is also shown. Sec. \ref{Fermion_correlation} is the main part of this work. We introduce the correlation functions of the Majorana CFT and establish the chiral boson-fermion correspondence at the level of the correlation function and the operators. In Sec. \ref{Numerical_Moore-Read}, applications of our method to a typical topologically ordered system, Moore-Read FQH states have been discussed. The interpretation of the conformal blocks and the corresponding wavefunctions can be seen more evidently compared with existing literature. Moreover, a method to generalize Laughlin's flux attachment argument to non-abelian particles is also discussed. It should be stressed that this generalized Laughlin's flux attachment argument implies the insufficiency of the existing framework for analysis for non-abelian anyons without considering the structure of the simple current extension. In Sec. \ref{generalization_simple_current}, general operator-state correspondence in a CFT with $Z_{N}$ simple current has been discussed. Sec. \ref{conclusion} is the concluding remark of this work, and we discuss the relationship between our method and those of existing literature and introduce open problems. In Appendix, we note technical or algebraic detail of the analysis in the manuscript and reltion between our formalism and existing ones. For the readers interested in the status or background knowledge of our formalism in this work, we note Appendix \ref{mathematical_literature}.

\section{Fusion rule of Ising and Majorana CFT}
\label{fusion_section}

First, let us introduce the chiral structure of the Ising CFT and Majorana CFT (For the readers interested in this aspect, we note a general review and textbook \cite{Ginsparg:1988ui,DiFrancesco:1997nk}). The Ising conformal field theory (CFT) which is frequently represented as the $M(3,4)$ minimal model has three primary fields, $\{ I ,\psi , \sigma\}$, with conformal dimension $h_{I}=0, h_{\psi}=1/2, h_{\sigma}=1/16$. $I$ corresponds to the identity operator, $\psi$ to the $Z_{2}$ simple current (or $Z_{2}$ symmetry generator), and $\sigma$ to the $Z_{2}$ chiral order (or spin) operator. They satisfy the following fusion rules:
\begin{align}
\psi\times \psi&= I, \\
\psi\times \sigma&= \sigma, \\
\sigma\times \sigma&=I+\psi.
\end{align}
Each fusion rule can be identified with the operator product expansions (OPEs) and their asymptotic behaviors of the fields. Because of its close relation to a one-dimensional quantum spin chain and two-dimensional statistical model, one can identify this chiral CFT as a bosonic CFT\cite{Dotsenko:1984nm,Dotsenko:1984ad}.
It is also a non-abelian CFT because of the absence of the inverse operator for $\sigma$. Hence, one can expect an exotic ``non-abelian" phenomenon in some related model in condensed matter physics, for example in the Moore-Read states which are celebrated fractional quantum Hall (FQH) systems\cite{Moore:1991ks}. It is remarkable that, in such systems, non-abelian anyons can be fundamental building blocks for topological quantum computations\cite{Mong_2014}.

However, there exists another aspect of this CFT with an abelian representation of the fusion rules (This is well known in high-energy physics, see \cite{Ginsparg:1988ui}, for example):
\begin{align}
\psi\times \psi&= I, \quad e\times m=\psi, \\
\psi\times e&= m, \quad\psi\times m= e, \\
e \times e&= m\times m =I.
\end{align}
This can be implemented by the identification $\sigma =(e+m)/\sqrt{2}$ where $e$ and $m$ are called semions with fermion parity even and odd and with conformal dimension $1/16$. This is a consequence of fermionization\cite{Ginsparg:1988ui,Runkel:2020zgg,Runkel:2022fzi}, because all non-trivial operators here have a Fermion parity. In addition, the fusion rule is now abelian, with $\psi, e, m$ as their own inverse operator.

It should be noted that the conformal dimension of semions is anomalous in the sense of generalized $Z_{2}$ Lieb-Shultz-Mattis anomaly classification, whereas they produce the $Z_{2}$ operation in the level of fusion rule. }There exist studies on the effect of the anomaly on the partition functions and the interpretation using t'Hooft anomaly\cite{Seiberg:2023cdc}. However, the interpretation of this anomalous dimension for the operators and the correlation functions has never been studied, as far as we know. We will clarify them in the following sections.

It may also be worth noting that the representation of chiral disorder operator $\mu =(e-m)/\sqrt{2}$ is a bosonic object but is difficult to notice from the original bosonic fusion rule (They satisfy the modified fusion rule $\psi\times\mu=-\mu$, $\mu\times \mu=1-\psi$. Appearance of this object can be seen in the recent study of generalized symmetry \cite{Li:2023mmw}). This implies the chiral Kramers-Wannier duality $\psi \rightarrow -\psi$ (and $m\rightarrow -m$). The above fermionized CFT is known as the Majorana CFT, appearing commonly in theoretical physics literature \cite{Ginsparg:1988ui}, and its fusion rule is nothing but Kitaev's double semion model\cite{Kitaev:2006lla}.

Formally, one can obtain the following fusion rule,
\begin{equation}
\sigma \times \mu=0.
\end{equation}
In other words, the observable containing both chiral order and disorder fields are outside of the usual bosonic chiral CFT. Related arguments can be seen in the studies of mixed state topological orders in the name premodular fusion category\cite{Wang:2023uoj,Ellison:2024svg,Kikuchi:2024ibt}.

In short, as we will formulate it more rigorously in the later sections, we have a chiral bosonic theory with primary operators, $\{ I, \psi , \sigma, \mu \}$ and the primary states labeled by $\{ I, \psi , \sigma \}$ (or $\{ I, \psi , \mu \}$ for the KW dual theory). From this expression, one can see that there exists one operator which does not have a corresponding state. On the other hand, the fermionic counterparts are $\{ I, \psi , e, m \}$ (conventionally one can introduce a notation $\{ I, -\psi , e, -m \}$ for its KW dual\cite{Hsieh:2020uwb,Fukusumi:2020irh,Fukusumi:2021zme}). One can see that the operators and states have a one-to-one correspondence in this fermionic theory. This is benefit of studying simple current extended CFTs. It should be emphasized that all of the operators can be identified only from their asymptotic behaviors at this stage.

As is well known in the literature \cite{Petkova:1988cy,Furlan:1989ra,Runkel:2020zgg}\footnote{For the readers interested in the categorified descriptions of group extension of full CFTs (or spherical fusion categories), we note recent works \cite{bischoff2021computingfusionrulesspherical,davydov2021braidedpicardgroupsgraded}.}, in the Ising CFT or Majorana CFT, there exist the set of bulk operators $\{ \psi, \overline{\psi}, \epsilon, \sigma_{\text{Bulk}},\mu_{\text{Bulk}}\}$ satisfying the following fusion rule:
\begin{align}
\epsilon&=\psi \overline{\psi},\quad \epsilon\times \epsilon=I \\
\sigma_{\text{Bulk}}\times \sigma_{\text{Bulk}}&=I+\epsilon, \mu_{\text{Bulk}}\times \mu_{\text{Bulk}}=I+\epsilon, \\
\sigma_{\text{Bulk}}\times \mu_{\text{Bulk}}&=\psi+\overline{\psi}, \\
\psi\times \sigma_{\text{Bulk}}&=\overline{\psi}\times \sigma_{\text{Bulk}}=\mu_{\text{Bulk}}, \\
\psi\times \mu_{\text{Bulk}}&=\overline{\psi}\times \mu_{\text{Bulk}}=\sigma_{\text{Bulk}},
\end{align}
where we have used the notations $\sigma_{\text{Bulk}}$ and $\mu_{\text{Bulk}}$ {to distinguish them from the chiral order field $\sigma$ and disorder field $\mu$, and $\psi$ ($\overline{\psi}$) is the chiral (antichiral) fermionic field, and $\epsilon$ is the energy field. The chiral and antichiral conformal dimensions of $\sigma_{\text{Bulk}}$ and $\mu_{\text{Bulk}}$ are both $1/16$ as we will demonstrate in the following discussions.

As is also wellk nown in the literature\cite{Petkova:1988cy,Furlan:1989ra,Runkel:2020zgg} (modern interpretations can be seen in the author's previous works \cite{Fukusumi:2022xxe,Fukusumi_2022_c}), the partition function of the Neveu-Schwartz (NS) and Ramond (R) sectors of Majorana CFT are,
\begin{align}
\text{NS}:& \ |\chi_{I}+\chi_{\psi}|^{2} \\
\text{R}:&\ 2|\chi_{\sigma}|^{2}
\label{Ramond-character}
\end{align}
where $\chi$ is the character labeled by the primary fields.
Only from this representation, the meaning of the prefactor ``2" in the R sector is not clear. To clarify this, let us consider the Hilbert space of fermionic theory. A naive application of the construction mimicking the method in \cite{Runkel:2020zgg} results in the following (extended) Hilbert space of R sector,
\begin{equation}
\mathcal{H}_{\text{R}}=(\mathcal{V}_{e}\oplus\mathcal{V}_{m})\otimes(\overline{\mathcal{V}}_{e}\oplus\overline{\mathcal{V}}_{m})
\end{equation}
where $\mathcal{V}$ is the Verma module labeled by the primary fields.
However, this results in the partition function
\begin{equation}
4|\chi_{\sigma}|^{2}.
\label{naive-decomposition}
\end{equation}
This naive decomposition Eq. \eqref{naive-decomposition} is different from the desired form Eq. \eqref{Ramond-character}. Hence, it is necessary to reduce the naive four-dimensional representation derived from the naive chiral-antichiral decomposition to a two-dimensional decomposition.} One needs to apply the Gliozzi-Scherk-Olive (GSO) projection $\{1\pm(-1)^{F+\overline{F}}\} /4$ to the Hilbert space of R sector corresponding to each parity where $F$ ($\overline{F}$) is the (anti)chiral fermion parity operator\cite{Gliozzi:1976qd}. For the readers interested in its interpretations in the lattice models, see recent works \cite{Hsieh:2020uwb,Seiberg:2023cdc} and the author's related work \cite{Fukusumi:2022xxe}, for example. For simplicity, let us take $\{ 1+(-1)^{F+\overline{F}}\} /4$ as GSO projection  to the fermion parity even sector.

In this case, one can obtain the Hilbert space (or ``physical Hilbert space") after the GSO projection as
\begin{equation}
\mathcal{H}_{f,\text{R}}=\mathcal{H}_{\sigma_{\text{Bulk}}}
\end{equation}
which can be identified as the conformal tower constructed from the following fermion parity even primary states under the identification $\sigma_{\text{Bulk}}=(e\overline{e}+m\overline{m})/\sqrt{2}$,
\begin{equation}
\frac{1}{\sqrt{2}}\left(|e\rangle \overline{|e\rangle}+|m\rangle \overline{|m\rangle}\right)=\frac{1}{\sqrt{2}}\left(|\sigma\rangle \overline{|\sigma\rangle}+|\mu\rangle \overline{|\mu\rangle}\right).
\end{equation}
The following sector constructed from the other GSOprojection $\{1-(-1)^{F+\overline{F}}\} /4$ corresponds to the fermion parity odd sector,
\begin{equation}
\mathcal{H}_{f',\text{R}}=\mathcal{H}_{\mu_{\text{Bulk}}}
\end{equation}
where one can identify the corresponding primary state as 
\begin{equation}
\frac{1}{\sqrt{2}}\left(|e\rangle \overline{|m\rangle}+|m\rangle \overline{|e\rangle}\right)=\frac{1}{\sqrt{2}}\left(|\sigma\rangle \overline{|\sigma\rangle}-|\mu\rangle \overline{|\mu\rangle}\right)
\end{equation}
by identifying $\mu_{\text{Bulk}}=(e\overline{m}+m\overline{e})/\sqrt{2}$.

Consequetly, one can obtain the two dimensional representation  (In this model, the sector $\sigma\overline{\mu}$ and $\mu\overline{\sigma}$ are projected out. One can see these sectors correspond to the dual of the physical Hilbert space. A similar (possibly the same) duality can be seen in \cite{Runkel:2020zgg}).
One can check that this expression, $\sigma_{\text{Bulk}}=(e\overline{e}+m\overline{m})/\sqrt{2}$  $\mu_{\text{Bulk}}=(e\overline{m}+m\overline{e})/\sqrt{2}$, by the chiral and antichiral fields reproduces the fusion rule of the bulk fields as we demonstrate in Appendix \ref{reconstruction}. It should be stressed that the unusual factor $1/\sqrt{2}$ is inevitable to apply the chiral antichiral decomposition of the theory, and this is a reason why the simple current extension of conformal field theory has not been solved in general. By replacing the GSO projection to the fermion parity even sector as $\{ 1-(-1)^{F+\overline{F}}\}/4$, one can interchange the order and disorder operator. This interchange corresponds to the parity shift operation in the work by Runkel and Watts \cite{Runkel:2020zgg} and appears in the work\cite{Seiberg:2023cdc} more recently. Related discussion about the counting of operators can be seen in \cite{Fukusumi:2022xxe,Fukusumi_2022_c,Lou:2020gfq}. Corresponding to these bulk sectors, one can associate the NS sector as Majorana edge modes and (fermionically) fixed boundary conditions as the R sector when studying boundary CFT and the corresponding $1+1$ dimensional quantum systems with open boundaries\cite{Runkel:2020zgg,Smith:2021luc, Weizmann,Fukusumi:2021zme}. For the R sector, one can distinguish them also by Kramers-Wannier (KW) $Z_{2}$ symmetry which only acts on the boundary\cite{Fukusumi:2020irh}, and this $Z_{2}$ and KW-$Z_{2}$ structure are closely related to the recent introduction of the categorical symmetry\cite{Kong:2014qka,Chatterjee:2022tyg}.

Before going to the next section, we note the ambiguity of the representation $\{ \sigma, \mu\}$ and $\{ e, m\}$ in the literature. Whereas $\{ e, m\}$ are fermionic primary fields distinguished by their fermionic parity, $\{ \sigma , \mu\}$ should be considered as bosonic operators distinguished by the sign of the operator $m$ field as $(e\pm m)/\sqrt{2}$. Unfortunately, this sign before $m$ field is also called parity, and these two parities might be confusing for the readers. This parity appears in $c=1$ CFT as the sign of the bosonic field $\text{cos}m\phi =(e^{im\phi}+e^{-im\phi})/2$ and $\text{sin}m\phi =(e^{im\phi}-e^{-im\phi})/2i$ respectively. Probably because of these two fermionic and bosonic parities, there has been some confusion between the usage of $\{ \sigma , \mu\}$ and $\{ e, m\}$ (or even $\{ \sigma_{\text{Bulk}} , \mu_{\text{Bulk}}\}$). This is the case even for several notable works and reviews. To some extent, this might be one of the sources of difficulties in calculating correlation functions and the corresponding Hilbert space of simple current extended CFTs.

\section{Asymptotic behaviors from fermionizations}
\label{Fermion_correlation}

Recently, it gradually becomes evident that the underlying CFT for the FQH system is a fermionic CFT ($Z_{N}$ simple current extended CFT more generally)\cite{Ino:1998by,Cappelli:2010jv,Schoutens:2015uia,Fuji_2017,Fukusumi_2022_c,Henderson:2023kjz}. In the previous works,  they have mainly tried to construct the corresponding partition function of CFT for the edge modes in the FQH systems. However, it is still unclear whether one can implement such a theory directly by constructing the multipoint correlation functions in a fermionic CFT and the corresponding wavefunctions in the FQH system. Hence it is also unclear whether one can interpret the asymptotic behaviors of correlation functions by using fermionic field theoretic representations. In short, in this section, we will demonstrate the fermionic interpretation of conformal blocks, which is much more concise and intuitive compared with the existing method. As we have mentioned in the previous sections, we mainly follow the discussions in the notable review \cite{Ginsparg:1988ui}, and compare the results with the other notable work by Ardonne and Sierra\cite{Ardonne:2010hj}. It is worth mentioning that the Kitaev's double semion fusion rule which captures attentions in recent studies on the fermionic topological order has already appeared in \cite{Ginsparg:1988ui}.

Here, we introduce a fermionic analog of the multipoint correlation function of Majorana CFT, which describes the asymptotic behavior of the general multipoint correlation function determined by Belavin-Polyakov-Zamolodchikov (BPZ) equation\cite{Belavin:1984vu}. For the readers interested in this aspect, we note a lecture note by Ribault \cite{Ribault:2016sla} (Also see Appendix \ref{CFT_structure} of this work).  It should be stressed that only asymptotic behaviors can be obtained in general as we demonstrate in Sec.\ref{exact_result} \footnote{In other words, the operator-state correspondence in the fermionic theory is broken in a strong sense, but there still exists a weaker version of operator-state correspondence as we will show.}. The procedures can be summarized as follows:

\begin{enumerate}
\item{Assuming the form of the multipoint correlation function of (non-anomalous) simple currents;}
\item{Assuming the mode expansion of simple currents and adding a twist to such mode expansion and identifying it to the insertion of semion;}
\item{Applying the contraction rule to the simple currents and defining the multipoint semion correlation function consistently.}
\end{enumerate}
Whereas we concentrate on the Majorana CFT, the above procedures can be applied to general CFTs with $Z_{N}$ simple current \cite{Blumenhagen:1990jv}\footnote{In the simple current extension, one can avoid the nilpotent fusion rule such as $\sigma \times \mu$ by introducing semions.}.

Firstly, we assume the following established form of CFT correlation function on a complex plane\cite{Ginsparg:1988ui,DiFrancesco:1997nk}
\begin{equation}
\langle \prod_{i=1}^{N} \psi (z_{i})\rangle=\text{Pf} \left[\frac{1}{z_{i}-z_{j}}\right] \label{MR_Pfaffian}
\end{equation}
where we have taken $N$ as an even integer and $\{ z_{i}\}_{i=1}^{N}$ is the complex variable.

Next, by mode expansion and normal ordering, we can calculate the following twisted correlation function which appears literature commonly,
\begin{equation}
\begin{split}
&\langle e(\infty) \prod_{i=1}^{N} \psi (z_{i}) e(0)\rangle=\langle m(\infty) \prod_{i=1}^{N} \psi (z_{i}) m(0)\rangle \\
&=\text{Pf} \left[\frac{1}{z_{i}-z_{j}}\left( \sqrt{\frac{z_{i}}{z_{j}}}+\sqrt{\frac{z_{j}}{z_{i}}}\right)\right].
\end{split}
\end{equation}
By applying the conformal transformation $z\rightarrow z+\omega$, and redefinition of the variable $z_{i}$, one can express it as
\begin{equation}
\begin{split}
&\langle e(\infty) \prod_{i=1}^{N} \psi (z_{i}) e(\omega)\rangle=\langle m(\infty) \prod_{i=1}^{N} \psi (z_{i}) m(\omega)\rangle \\
&=\text{Pf} \left[\frac{1}{z_{i}-z_{j}}\left( \sqrt{\frac{z_{i}-\omega}{z_{j}-\omega}}+\sqrt{\frac{z_{j}-\omega}{z_{i}-\omega}}\right)\right].
\end{split}
\end{equation}

Then, one would introduce the following (abelianized) multipoint correlation function,
\begin{equation}
\langle \prod_{i=1}^{M} e (\omega_{i})\rangle .
\end{equation}

However, the direct calculation of this correlation function contains subtleties as we will demonstrate in Sec.\ref{exact_result} (As far as we know, this problem has never been noticed in the literature.). Instead, we introduce the following abelianized correlation function which contains sufficient information to define the asymptotic behavior of the above correlation function
\begin{equation}
\prod_{i=1}^{M/2}\langle e (\omega_{g(i)_{1}}) e (\omega_{g(i)_{2}})\rangle=\prod_{i=1}^{M/2}(\omega_{g(i)_{1}}-\omega_{g(i)_{2}})^{-1/8}
\end{equation}
where $g$ is a partition which divide $M$ (an even integer) semion into two groups and each $g(i)_{k}$, $k=1,2$ are paired.

Also one can define the following correlation function,
\begin{equation}
\begin{split}
&\langle \prod_{i=1}^{N} \psi (z_{i}) \prod_{i'=1}^{M}e(\omega_{i'})\rangle=\langle \prod_{i=1}^{N} \psi (z_{i}) \prod_{i'=1}^{M}m(\omega_{i'})\rangle \\
&\xrightarrow{g} \text{Pf} [\frac{1}{z_{i}-z_{j}}\prod_{k=1}^{M/2}\{ \sqrt{\frac{z_{i}-\omega_{g(k)_{1}}}{z_{j}-\omega_{g_(k)_{1}}}}\sqrt{\frac{z_{j}-\omega_{g(k)_{2}}}{z_{i}-\omega_{g_(k)_{2}}}} \\
&+\sqrt{\frac{z_{j}-\omega_{g(k)_{1}}}{z_{i}-\omega_{g_(k)_{1}}}}\sqrt{\frac{z_{i}-\omega_{g(k)_{2}}}{z_{j}-\omega_{g_(k)_{2}}}} \} ] \\
&\times\prod_{k=1}^{M/2}(\omega_{g(k)_{1}}-\omega_{g(k)_{2}})^{-1/8}
\end{split}
\label{correlation_fermion}
\end{equation}
for $M$ even, where  $\xrightarrow{g}$ corresponds to introduction of the partition $g$. They are the fermionic analog of the ``pairing" of quasihole introduced in studying fractional quantum Hall effect (FQHE) \cite{Moore:1991ks,Milovanovic:1996nj} (By replacing $g(i)_{j}$ in this work to permutation $\sigma (2i+j-2)$, one can obtain the standard notation in \cite{Moore:1991ks,Milovanovic:1996nj}. The paring corresponds to ``condensation" of semionic objects in the recent literature\cite{Ellison:2024svg,Wang:2023uoj}.). One can obtain the same form of the correlation function for odd $M$ by adding $e(\infty)$ or $m(\infty)$. 

As one may have already noticed, the righthand side of the above equation only contains the power $(\omega_{i}-\omega_{j})^{-1/8}$. This corresponds to the identity channel of fusion rule $\sigma \times \sigma =I +\psi$. Hence to obtain the non-abelian contributions of $\sigma$, one has to introduce correlation functions containing both $e$ and $m$. Eq. \eqref{correlation_fermion} seems to correspond to the (bulk) multiple quasihole wavefunctions in the existing literature, but they shouldn't be identified with the multipoint correlation functions constructed from $\sigma$. For the operator $\sigma$, one needs to calculate (or define) correlation functions containing both $e$ and $m$.

For this purpose, let us assume the following operator product expansion (OPE) inferred from the fusion rule,
\begin{equation}
\psi(z) e(\omega)\sim (z-\omega)^{-1/2} m(\omega).
\end{equation}
In other words, one can define the $m$ operator as,
\begin{equation}
m(\omega)= \lim_{z\rightarrow \omega} (z-\omega)^{1/2}\psi(z) e(\omega).
\end{equation}
We use the above OPE to derive the correlation function containing both $e$ and $m$, starting from the proposed correlation function only containing $e$ and $\psi$.

As a simplest calculation, one can check,

\begin{equation}
\begin{split}
&\langle e(\omega_{1})m(\omega_{2})\psi (z)\rangle \\
&= \frac{1}{(\omega_{1}-\omega_{2})^{-3/8}(z-\omega_{1})^{1/2}(z-\omega_{2})^{1/2}}
\end{split}
\end{equation}

\begin{equation}
\begin{split}
&\langle e(\omega_{1})e(\omega_{2})m(\omega_{3})m(\omega_{4})\rangle \\
&\xrightarrow{(1,2)(3,4)} \langle e(\omega_{1})e(\omega_{2})\rangle \langle m(\omega_{3})m(\omega_{4})\rangle\\
&= (\omega_{1}-\omega_{2})^{-1/8} (\omega_{3}-\omega_{4})^{-1/8} \\
&\times\left( \sqrt{\frac{\omega_{1}-\omega_{3}}{\omega_{2}-\omega_{3}}\frac{\omega_{2}-\omega_{4}}{\omega_{1}-\omega_{4}}}+\sqrt{\frac{\omega_{1}-\omega_{4}}{\omega_{2}-\omega_{4}}\frac{\omega_{2}-\omega_{3}}{\omega_{1}-\omega_{3}}}\right)
\end{split}
\end{equation}
and
\begin{equation}
\begin{split}
&\langle e(\omega_{1})m(\omega_{2})e(\omega_{3})m(\omega_{4})\rangle \\
&\xrightarrow{(1,2)(3,4)} \langle e(\omega_{1})m(\omega_{2})\rangle \langle e(\omega_{3})m(\omega_{4})\rangle\\
&= (\omega_{1}-\omega_{2})^{3/8} (\omega_{3}-\omega_{4})^{3/8} (\omega_{1}-\omega_{4})^{-1/2} (\omega_{2}-\omega_{3})^{-1/2}
\end{split}
\end{equation}
where we have taken the partition $(1,2) (3,4)$ for simplicity.

In general, one can represent the following series of correlation functions,
\begin{equation}
\begin{split}
&\langle \prod_{i=1}^{M} e(\omega_{i}) \prod_{i'=1}^{M'} m(\omega'_{i'}) \prod_{j=1}^{N} \psi(z_{j})\rangle \\
&= \lim_{\{z_{N+i'}\rightarrow \omega'_{i'}\}^{M'}_{i'=1}}\langle \prod_{i=1}^{M} e(\omega_{i}) \prod_{i'=1}^{M'} e(\omega'_{i'})
\prod_{j=1}^{N+M'} \psi(z_{j})\rangle \\
&\times\prod_{i'=1}^{M'} (z_{N+i'}-\omega'_{i'})^{1/2}.
\end{split}
\label{correlation_generic}
\end{equation}
where we have taken the integers $M$, $M'$, $N$ such as $M+M'$ and $M'+N$ become even, and introduced $\{ \psi (z_{j})\}_{j=N+1}^{N+M'}$ to represent the effect of $\{m(\omega'_{i'})\}_{i'=1}^{M'}$. Hence, one can insert any number of $e$, $m$, $\psi$, $\sigma$, $\mu$, as one wants and place them to $\infty$. However, it should be stressed this may not result in exotic states when considering bulk-edge correspondence. In this section, we only consider the parity even part, but taking some variables to $\infty$, one can obtain parity odd part in the FQH systems.

In the condensed matter physicists community studying FQH systems, the energy equivalence between this electron odd sector and even sector is called supersymmetry (``SUSY")\cite{Yang_2012,Yang_2013,Gromov:2019cgu,Sagi:2016slk,Ma:2021dua,Nguyen:2022khy}. However, this is usually called $Z_{2}$ symmetry in other research communities, including statistical mechanics, high energy physics, and mathematical physics. Moreover, this seems different from the more conventional supersymmetry in the string theory that implies a correspondence between $Z_{2}$ charged and uncharged sector\cite{Gliozzi:1976qd,Fukusumi:2022xxe} (For the audience interested in the historical aspect, we cite the legacy of Olive\cite{https://doi.org/10.48550/arxiv.2009.05849}). Here, supersymmetry is a correspondence between $\{ I, \psi\}$ and $\{ e,m\}$, whereas ``SUSY" in the contemporary literature is $\{ I, e\}$ and $\{ \psi,m\}$. The latter is usually called Ising $Z_{2}$ symmetry and one can see this even in the bosonic partition function of the Ising model (In this sense, the usual Kitaev chain and Ising chain and all $Z_{2}$ symmetric system have ``SUSY"). Alternatively, one can analyze (bosonic or fermionic) superconformal field theory based on spin $3/2$ supercurrent, but, in general, this is also different from the fermionic theory defined by the $Z_{2}$ simple current terminologically. Moreover, wavefunctions of FQHE apparently do not contain Grassmann variables which should appear in studying the correlation functions of supersymmetric conformal field theories. The related terminological problem in fermionic CFT has been discussed in \cite{Runkel:2020zgg}. 

\subsection{Comparison with the exact results}
\label{exact_result}

Here we review the exact form of the multipoint correlation function of the Ising model and propose their renewed interpretation based on the fermionization. The most relevant reference for the construction of the chiral correlation function is the work by Ardonne and Sierra \cite{Ardonne:2010hj}. Theoretically, the structures of the correlation of the Ising model have been determined by mapping the relation between the doubled Ising model and the free boson model (or corresponding Dirac fermion model)\cite{DiFrancesco:1987ez}. In this mapping, they focused on the correspondence between non-abelian objects in the free boson (or abelian) theory and the corresponding order and disorder operators in the Ising CFT. Several earlier attempts to represent the chiral correlation functions can be seen in the studies of Moore-Read fractional quantum Hall states\cite{Arovas:1984qr,Milovanovic:1996nj,Fendley_2006,Fendley_2007}. ``Lattice" construction of the Moore-Read states can be seen in \cite{Glasser:2015fla} based on the results of \cite{Ardonne:2010hj}.

As in the previous section, it may be reasonable to identify the correlation function corresponding to the fermionic correlation function, $\prod_{i} e(\omega_{i})\prod_{i}\psi(z_{i})$ or $\prod_{i} m(\omega_{i})\prod_{i}\psi(z_{i})$, but this seems quite difficult or impossible as we will demonstrate. By considering asymptotic behaviors of the solutions of the BPZ equations, we can identify the (chiral) conformal block of correlation function of $\prod_{i} (e(\omega_{2i-1})e(\omega_{2i})+m(\omega_{2i-1})m(\omega_{2i}))\prod_{i}\psi(z_{i})$ as,
\begin{equation}
\begin{split}
&\langle\prod_{i} (e(\omega_{2i-1})e(\omega_{2i})+m(\omega_{2i-1})m(\omega_{2i}))\prod_{i}\psi(z_{i})\rangle \\
&=\prod_{i=1}^{M/2}(\omega_{2i-1}-\omega_{2i})^{-1/8}\prod_{i=1}^{M}\prod_{j=1}^{N}(\omega_{i}-z_{j})^{-1/2} \\
&\times A_{N} (\{ x_{i,j} \})^{-1/2} \\
&\times\sum_{t_{1}=1, t_{2}...t_{N/2}=\pm1}\prod_{1\le i,j \le N/2} (1-x_{i,j})^{t_{i}t_{j}/4} \Psi_{\{t_{i}\}}(\{ \omega_{i}\}, \{ z_{i}\})
\end{split}
\label{Majorana_conformal_block}
\end{equation}
where we have introduced the cross ratio,
\begin{equation}
x_{i,j}=\frac{(\omega_{2i-1}-\omega_{2i})(\omega_{2j-1}-\omega_{2j})}{(\omega_{2i-1}-\omega_{2j})(\omega_{2j-1}-\omega_{2i})}
\end{equation}
and the following functions,
\begin{equation}
A_{N} (\{ x_{i,j} \})=\sum_{t_{1}=1, t_{2}...t_{N/2}=\pm1}\prod_{1\le i,j \le N/2} (1-x_{i,j})^{t_{i}t_{j}/4}
\label{aspect_factor}
\end{equation}

\begin{equation}
\begin{split}
&\Psi_{\{t_{i}\}}(\{ \omega_{i}\}, \{ z_{i}\}) \\
&=\text{Pf}\frac{\prod_{i}(\omega_{2i-\frac{t_{i}-1}{2}}-z_{j})(\omega_{2i+\frac{t_{i}-1}{2}}-z_{k})+(j\leftrightarrow k)}{z_{j}-z_{k}}
\end{split}
\end{equation}

For simplicity, let us check the form of $\langle (ee+mm)(ee+mm) \rangle$. The exact expression is,
\begin{equation}
\begin{split}
&\langle (e(\omega_{1})e(\omega_{2})+m(\omega_{1})m(\omega_{2}))(e(\omega_{3})e(\omega_{4})+m(\omega_{3})m(\omega_{4}))\rangle \\
&=(\omega_{1}-\omega_{2})^{-1/8}(\omega_{3}-\omega_{4})^{-1/8}\sqrt{(1-x)^{1/4}+(1-x)^{-1/4}}
\end{split}
\label{block1}
\end{equation}
where we have introduced the aspect ratio $x=\{ (\omega_{1}-\omega_{2})(\omega_{3}-\omega_{4})\} /\{ (\omega_{1}-\omega_{4})(\omega_{2}-\omega_{3})\} $ to simplify the notation. Here, we check the consistency of this expression by considering the asymptotic behaviors of the righthand side of Eq.\eqref{block1} (To simplify the discussion, we fix the operator at $\omega_{1}$ as $e$). When considering the asymptotic behavior $x\sim 0$, one can obtain the righthand side of Eq.\eqref{block1} as $ (\omega_{1}-\omega_{2})^{-1/8}(\omega_{3}-\omega_{4})^{-1/8} $. This corresponds to $\langle e(\omega_{1})e(\omega_{2})\rangle\langle e(\omega_{3})e(\omega_{4})\rangle$ or $\langle e(\omega_{1})e(\omega_{2})\rangle\langle m(\omega_{3})m(\omega_{4})\rangle$ in the previous section. In this representation, one can see the asmptotics $x\sim 0$ turns to $\xrightarrow{(1,2)(3,4)}$ in Eq.\eqref{correlation_fermion}. In a similar way, we can identify the asmptotic behavior $x\sim 1, -\infty$ of this exact four point function as
 $\langle e(\omega_{1})e(\omega_{3})\rangle\langle e(\omega_{2})e(\omega_{4})\rangle$, $\langle e(\omega_{1})m(\omega_{3})\rangle\langle e(\omega_{2})m(\omega_{4})\rangle$ and $\langle e(\omega_{1})e(\omega_{4})\rangle\langle e(\omega_{2})e(\omega_{3})\rangle$, $\langle e(\omega_{1})m(\omega_{4})\rangle\langle e(\omega_{2})m(\omega_{3})\rangle$ respectively. One can see the identifications of the operations between $x\sim 1$ as $\xrightarrow{(1,3)(2,4)}$, and $x\sim -\infty$ as $\xrightarrow{(1,4)(2,3)}$.
 
As one may have already noticed, the encircling process between $\omega_{2}$ and $\omega_{3}$ changes the form of the correlation function and this corresponds to the insertion of the fermionic field $\psi$ at the position $\omega_{2}$ and $\omega_{3}$ combined with a phase factor. 

One can consider the same analysis for the other four-point block, 
\begin{equation}
\begin{split}
&\langle (e(\omega_{1})m(\omega_{2})+m(\omega_{1})e(\omega_{2})) (e(\omega_{3})m(\omega_{4})+m(\omega_{3})e(\omega_{4}))\rangle \\
&=(\omega_{1}-\omega_{2})^{-1/8}(\omega_{3}-\omega_{4})^{-1/8}\sqrt{(1-x)^{1/4}-(1-x)^{-1/4}}.
\end{split}
\label{block2}
\end{equation} 
However, it is impossible to obtain the exact correlation function corresponding to $\langle eeee\rangle$ from a linear combination of these two conformal blocks, Eq.\eqref{block1} and Eq. \eqref{block2}, for example. In this sense, the operator-state correspondence in the Majorana CFT is broken.

The most striking point of this representation of correlation function is one can identify this correlation function as Schottky double of bulk correlation function. The doubling trick of BCFT or Schottky double can be understood as a mapping of a antichiral field $\overline{\phi}$ to a chiral field $\phi$. For simplicity, we take the map as  $\overline{\phi}_{\alpha}$ to a chiral field $\phi_{\alpha}$ and denote 
\begin{equation}
\{\overline{\phi_{\alpha}}\}_{\text{SD}}=\phi_{\alpha}. 
\end{equation}
We also note that the Schottky double or the doubling trick\cite{PhysRevLett.54.1091,Cardy:1986gw} has been revisited recently in the name of ancillary CFT(ACFT)\cite{Nishioka:2022ook} (For the readers interested in this aspect, see the review \cite{Schweigert:2000ix} and reference therein. We also note \cite{Kawai:2002vd} as a typical application).

In short, one can write down the general correlation function Eq. \eqref{Majorana_conformal_block} as (For the readers interested in the detailed calculation, see the discussion in Appendix \ref{algebra_OPE}),
\begin{equation}
\begin{split}
&\langle\prod_{i=1}^{M/2} \frac{e(\omega_{2i-1})e(\omega_{2i})+m(\omega_{2i-1})m(\omega_{2i})}{\sqrt{2}}\prod_{i=1}^{N}\psi(z_{i})\rangle \\
&=\langle\prod_{i=1}^{M/2} \{ \sigma_{\text{Bulk}}(\omega_{2i-1}, \omega_{2i})\}_{\text{SD}}\prod_{i=1}^{N}\psi(z_{i})\rangle
\end{split}
\end{equation}
By replacing $\sigma_{\text{Bulk}}$ with $\mu_{\text{Bulk}}$, this shifts the fermionic parity of the pair appearing in each position and one can obtain the general form of conformal blocks extensively studied in \cite{Ardonne:2010hj}\footnote{The diagrammatic decomposition of the sectors of $\sigma \times \sigma$ is also useful for the calculations, but we would like to stress that the naive introduction of the object $\sigma \overline{\sigma}$ (or $\sigma \otimes \overline{\sigma}$) leads to confusing representation in the context of simple current extension. Because of the algebraic relation $\psi \times (\sigma \overline{\sigma})=\sigma \overline{\sigma}$, introduction of the disorder field becomes difficult only by the parity shift operation\cite{Runkel:2020zgg,Hsieh:2020uwb}. Related discussions can be seen in \cite{Wen:2024udn}.}. Hence we propose a modified or weak operator-state correspondence: \emph{``the chiral CFT correlation function can be constructed by Schottky double of bulk CFT. For $Z_{2}$ extension of the Ising or Majorana fermion case, $\{I,\psi, \{ \sigma_{\text{Bulk}}\}_{\text{SD}},\{\mu_{\text{Bulk}}\}_{\text{SD}} \}$ and their descendants form a complete set".}

It may be worth noting the boson-fermion correspondence and its chiral extension at this stage. By using the relation, $\sigma_{\text{Bulk}}=(\sigma \overline{\sigma}+\mu \overline{\mu})/\sqrt{2}$, $\mu_{\text{Bulk}}=(\sigma \overline{\sigma}-\mu \overline{\mu})/\sqrt{2}$, one can easily observe the correlation functions can be constructed from the other basis of operators $\{I,\psi, \{ \sigma \overline{\sigma}\}_{\text{SD}}, \{\mu \overline{\mu}\}_{\text{SD}} \}$ or $\{I,\psi, \sigma\sigma, \mu \mu \}$ after SD. This may be surprising for the readers familiar with calculations of chiral correlation functions, because usually it depends on the detailed representation in formalisms like the Dotsenko-Fateev integral or vertex operator algebra. In contrast, here the form of correlation functions has been determined uniquely  by introducing the semions (or by \emph{extending} $\sigma$ to $e,m$) and this property is useful for further studies based on Matrix-Product-State (MPS) or Tensor-Network formalisms. Our formalism can be thought of as a generalization of the works \cite{Fuchs:2002cm,Fuchs:2004dz,Vanhove:2018wlb} with emphasis on their algebraic structures. For the readers interested in the theoretical (or mathematical) backgrouds, we summarized the relation between our formalism and the existing ones in Appendix \ref{mathematical_literature}.

It may be remarkable that a conformal block itself has not been considered as a ``physical'' quantity in considering statistical mechanical models, usually (Similar observation can be seen in the recent proposal on ACFT \cite{Nishioka:2022ook}). Hence, it is necessary to take a linear combination of these conformal blocks when considering an observable (or the partition function) of a lattice model. In the existing literature, the reason why the conformal block becomes ``unphysical" has been mysterious, but we have identified the conformal blocks as the fermionic correlation functions. In other words, in a class of models, the mismatch of the representations between bosonic CFT corresponding to the lattice observables and the fermionic CFT of the conformal block resolves this puzzle.  For example, related observations (corresponding to the correlation function in the chiral bosonic CFT) can be seen in the studies of multiple Schramm-Loewner evolution or crossing probabilities in lattice models\cite{Bauer:2005jt,Gori:2017cyq,Gori:2018gqx,Fukusumi:2020onf}. In this sense, the introduction of ``degeneracies'' coming from the conformal blocks is confusing, but we have clarified its meaning by establishing the correspondence to bulk correlation function including disorder operator and we name this correspondence as chiral CFT/ doubled CFT correspondence (CCFT/DCFT). Similar statements, like CFT/TQFT correspondence, have been proposed in the literature, but our expression clarifies its relation to fermionization in a concrete way. For example, one can relate the degeneracies of $2M$ quasihole states wavefunction of the Moore-Read states to the choice of the $M$ order or disorder operators, and one can read off the degeneracies as $2^{M-1}$, by considering their fermion parity.

\section{Bulk-edge correspondence and application to Moore-Read states}
\label{Numerical_Moore-Read}

The most significant point of the bulk-edge correspondence in the FQHE system is the correspondence between the labeling of the edge quantum states in FQH systems as quantum states (at infinity) in CFT. Hence, the edge states in Moore-Read states should be labeled by $I$, $\psi$, $e$, $m$, or $I$, $\psi$, $\sigma$, $\mu$, and their descendants coupled with $U(1)$ part. The existing literature seems to suggest $I$, $\psi$, and $\sigma$, is a natural set of labels when considering fermionic chains and the corresponding CFT, but this might be different in the FQH system. Historically this problem in labeling edge states has been pointed out in \cite{Milovanovic:1996nj} in relation to zero modes of FQH systems.

It should be stressed that after the invertiblization, one can see all of the states living in the edge and the operator living in the bulk should be paired (More mathematically, this is a kind of Atiyah-Singer index theorem). This is a charge (and parity) neutrality condition and one should count either the operator in the bulk or states at the edge to specify each situation.

Moreover, in the FQH system, one has to attach flux (or $U(1)$ CFT) to each primary field in CFT to obtain single-valued wavefunctions. Corresponding to the $Z_{2}$ simple charge of a CFT field, the flux becomes (usual) integer flux or (unusual) fractional flux. Especially in the fermionic case, this flux becomes an integer flux or half flux. For the operator $\{ I, \psi \}$, the attached flux becomes integer flux. It becomes a half-integer flux for the operator $\{ e, m \}$.

\subsection{Moore-Read wave function}
Here we demonstrate several sets of wavefunctions constructed from the BPZ equation exactly to become the zero energy state of the lowest Landau level of the microscopic model introduced in \cite{Wan2007FractionalQH}. This result shows the consistency of our method of constructing correlation functions of chiral CFT and the corresponding topological order. In other words, if our interpretation is invalid, it becomes necessary to test the fundamental structures of CFT/TQFT correspondence or the definition of chiral CFT itself.

One can write down the wave functions of the bosonic Moore-Read state and its quasihole excitations by combining the correlation functions of fermionic CFT with those of $U(1)$ chiral CFT (there exists the fermionic Moore-Read states, but let us concentrate on bosonic one for simplicity). The correlation function for the bosonic field is
\begin{equation}
\langle e^{i \phi(z_1)} \cdots e^{i \phi(z_N)}\rangle =\prod_{i<j}(z_i-z_j) \label{Jastrow_factor}
\end{equation}
By multiplying it with Eq.$\,$(\ref{MR_Pfaffian}), one can write down the bosonic Moore-Read wavefunction$\,$\cite{Moore:1991ks}
\begin{equation}
\Psi_{\text{MR}}=\text{Pf}[\frac{1}{z_i-z_j}]\prod_{i<j}(z_i-z_j).
\end{equation}
Thus, one can identify the operator $\psi(z)e^{i \phi(z)}$ as an electron operator. The Moore-Read wavefunction for fermions can be written down directly by multiplying $\Psi_{\text{MR}}$ with a Jastrow factor Eq.$\,$(\ref{Jastrow_factor}). Since $\Psi_{\text{MR}}$ vanishes when any three particles are at the same position, one can construct a parent Hamiltonian $\hat{H}_{\text{MR}}$ based on its clustering property so that $\hat{H}_{\text{MR}}\Psi_{\text{MR}}=0$, and the result is$\,$\cite{simon_generalized_2007}:
\begin{equation}
\hat{H}_{\text{MR}} = \sum_{i<j<k} \delta(\bm{r}_{i}-\bm{r}_{j})\delta(\bm{r}_{i}-\bm{r}_{k})\delta(\bm{r}_{j}-\bm{r}_{k})
\end{equation}
with $\delta(\bm{r})$ being a two-dimensional Dirac delta function. $\hat{H}_{\text{MR}}$ will punish a finite energy when any three particles are at the same position.

Similarly, one can identify $e(\omega)e^{i \phi(\omega)/2}$ as the quasihole operator and write down various quasihole wave functions for an even number of electrons by multiplying Eq.$\,$(\ref{correlation_fermion}) with the bosonic part given by
\begin{gather}
\langle \prod_{i'=1}^{M} e^{i \phi( \omega_{i'} )/2} \prod_{i}^{N} e^{i \phi(z_{i})} \rangle \label{correlation_boson} \\
=\prod_{i'<j'}(\omega_{i'}-\omega_{j'})^{1/4} \prod_{i}\prod_{i'}(z_{i}-\omega_{i'})^{1/2} \prod_{i<j} (z_i -z_j). \nonumber
\end{gather}
The explicit expression of the quasihole wave function is
\begin{gather}
\Psi_{\text{MR}}^{\text{QH}}=\text{Pf} \left[\frac{\prod_{k=1}^{M/2} \left[ (z_{i}-\omega_{g(k)_{1}})(z_{j}-\omega_{g(k)_{2}}) +(i \leftrightarrow j) \right]}{z_{i}-z_{j}} \right] \nonumber \\
\times\prod_{k=1}^{M/2}(\omega_{g(k)_{1}}-\omega_{g(k)_{2}})^{-1/8} \prod_{i'<j'}(\omega_{i'}-\omega_{j'})^{1/4} \prod_{i<j} (z_i -z_j).
\label{Quaishole_WF}
\end{gather}
By acting $\hat{H}_{\text{MR}}$ on $\Psi_{\text{MR}}^{\text{QH}}$, it is direct to show $\hat{H}_{\text{MR}}\Psi_{\text{MR}}^{\text{QH}}=0$, which follows from the property of the delta function. Thus, they are also zero-energy states of $\hat{H}_{\text{MR}}$. Moreover, one can write down various quasihole wave functions for a generic number of electrons by multiplying Eq.$\,$(\ref{correlation_generic}) and Eq.$\,$(\ref{correlation_boson}). It can be shown they are also zero-energy states of $\hat{H}_{\text{MR}}$ by using the same method.

The linear independent quasihole wavefunctions corresponding to different conformal blocks can also be written down by multiplying them with Eq.$\,$(\ref{correlation_boson}). Since they are linear combinations of quasihole wave functions in Eq.$\,$(\ref{Quaishole_WF}), they are also zero-energy states of $\hat{H}_{\text{MR}}$(Hence, we can drop off complicated factors depending on the aspect ratio, such as Eq. \eqref{aspect_factor}, in the above discussion). In this way, we can construct $2^{M/2-1}$ linear independent degenerate states for $M$ quasiholes. Then, one can use them to analyze the property of Moore-Read quasiholes in the FQH system. Since the braiding between Moore-Read quasiholes is non-abelian, it is proposed that they can be used to realize the topological quantum computation and attract a lot of attention$\,$\cite{nayak_non-abelian_2008}.

\subsection{Toward second quantization of non-abelian objects}

Recently, a second quantized formulation of quasihole operators has been proposed in \cite{Mazaheri_2015}. This method is potentially useful for establishing both an intuitive and quantitive understanding of FQHE based on the microscopic model. This has been a central concern in condensed matter theorist communities\cite{Wan2007FractionalQH,simon_generalized_2007}, as we have explained in the previous subsection. However, the generalization of this method to non-abelian particles is still in progress. Moreover, the interpretation of degeneracies appearing in FQH systems is unclear in this method. Our method to formulate wavefunctions by using SD may give a clear understanding of this puzzle. First of all, the notion of the single quasihole operator attached to half-flux quantum is subtle, because of the ambiguity of $e, m$ or $\sigma, \mu$. Hence it is necessary to introduce paired operator $\sigma_{\text{Bulk}} (\omega_{1},\omega_{2})$ and $\mu_{\text{Bulk}} (\omega_{1}, \omega_{2})$. For each paired operator, one can see the fermionic parity $0$ or $1$ and one can represent the corresponding wavefunction by specifying the fermion parity of each pair. One can represent
\begin{equation}
\langle\prod_{i}^{M/2}\{\sigma_{\text{Bulk}}(\omega_{2i-1},\omega_{2i})\}_{\text{SD}}\prod_{j}^{N}\psi(z_{j})\rangle\rightarrow00...0= 0^{M/2}
\end{equation}
\begin{equation}
\langle\prod_{i}^{M/2}\{\mu_{\text{Bulk}}(\omega_{2i-1},\omega_{2i})\}_{\text{SD}} \prod_{j}^{N}\psi(z_{j})\rangle \rightarrow11...1= 1^{M/2}
\end{equation}
and so on. This representation gives a concise understanding of diagrammatic representation of the conformal blocks in \cite{Ardonne:2010hj}, and can be generalized to more general topological orders.

Consequently, our formulation suggests the necessity to introduce $\{ \{\sigma_{\text{Bulk}}\}_{\text{SD}},\{\mu_{\text{Bulk}}\}_{\text{SD}}\}$ or $\{ \sigma \sigma, \mu\mu\}$ which are indecomposable to one-point objects. This expression can give the generalization of Laughlin's argument to these non-abelian anyons. Moreover, the indecomposability in these objects implies entanglement at the level of operators and may signal (or even clarify) the usefulness of non-abelian anyons for quantum information processing. Further identifications of these paired non-abelian objects in the microscopic models or lattice systems may be fundamentally important in this direction (even for the experimental realizations).

\section{Toward application for more general simple current extended CFT and the corresponding topological order}
\label{generalization_simple_current}
In this section, we propose a possible application of our method to a general simple current extended CFT\cite{Gato-Rivera:1990lxi,Gato-Rivera:1991bcq,Gato-Rivera:1991bqv,Fuchs:1994sq,Fuchs:1996rq,Fuchs:1996dd,Fuchs:2004dz}. As one of the authors has emphasized in \cite{Fukusumi_2022_c}, one can apply a phenomenological understanding of quark confinement to such simple current extended models. Hence the untwisted parts can be interpreted as the bulk local operators and the twisted parts can be interpreted as bulk nonlocal operators, by using the particle physics basis (or lattice analog of parafermionized representation). In this language, it should be worth noting that the definition of locality has been changed compared with the corresponding bosonic basis. For simplicity, let us consider anomaly free case in \cite{Fukusumi_2022_c,Gato-Rivera:1990lxi,Gato-Rivera:1991bcq,Gato-Rivera:1991bqv}, corresponding to Lieb-Shultz-Mattis anomaly-free theory (such as $SU(N)_{KN}$ Wess-Zumino-Witten model), and introduce the following bosonic partition function,
\begin{equation}
Z_{M}=\sum_{i,p} \chi_{i,p}\overline{\chi}_{\overline{i},-p}+\sum_{a} |\chi_{a}|^{2},
\end{equation}
where $i$ and $a$ are labeling the $Z_{N}$ invariant and noninvariant states correspondingly, $p$ is labeling the $Z_{N}$ partity of them and the $Z_{N}$ charge conjugation for an operator $\phi_{\alpha}$ is $\phi_{\overline{\alpha}}$. (For simplicity, we have introduced the bosonic theory with $p+\overline{p}=0$. There can exist the other choice $p-\overline{p}=0$.) This representation distinguishing $Z_{N}$ invariant and noninvariant sectors has captured the attention of the theoretical physicists recently in \cite{Numasawa:2017crf,Kikuchi:2019ytf,Li:2022drc}. 

In this section, we denote the bulk fields as $\Phi$ and the chiral (antichiral) fields as $\phi$ ($\overline{\phi}$) following the notations in \cite{Fukusumi:2024cnl}. After introducing the semion $\phi_{a}=(\sum \phi_{a,p})/\sqrt{N}$, the partition function becomes,
\begin{equation}
Z_{PM,Q}=\sum_{i\in \{Q_{J}(i)=Q\}} |\sum_{p}\chi_{i,p}|^{2}+N\sum_{a\in \{Q_{J}(a)=Q\}} |\chi_{a}|^{2}
\end{equation}
where $J$ is the generator of $Z_{N}$ symmetry called $Z_{N}$ simple current and $Q_{J}(\alpha)=h_{J}+h_{\alpha}-h_{J\times\alpha}$, and $h$ is the (chiral) conformal dimension labeled by $i$, $p$ and so on.

Applying our method to this case, one can observe the appearance of bulk disorder operator for the $Z_{N}$ semion sectors and this results in the ambiguity of operator-state correspondence of chiral CFT. An easiest choice of the bulk local operator is,
\begin{align}
\Phi_{i, p,\overline{p}}&=\phi_{i,p}\overline{\phi}_{\overline{i},\overline{p}} \\
\Phi_{a, 0}&=\sum_{p'}\phi_{a,p'}\overline{\phi}_{\overline{a},-p'} /\sqrt{N}
\end{align}
As in the Majorana fermion case, we expect the above operators to have a local expression in the (bosonic) lattice model.
The bulk disorder operators can be identified with the space that has already been projected out by GSO projection,
\begin{equation}
\Phi_{a, p}=\sum_{p'}\phi_{a,p'+p}\overline{\phi}_{\overline{a},-p'} /\sqrt{N}.
\end{equation}
where we have taken GSO projection for the order fields as $\sum_{k=0}^{N-1} e^{\frac{2\pi ik}{N}(F+\overline{F})}/N^{2}$ and disorder fields as $\sum_{k=0}^{N-1} e^{\frac{2\pi ik}{N}(F+\overline{F}-p)}/N^{2}$ where $F$ ($\overline{F}$) is the chiral (antichiral) $Z_{N}$ parity operator. As has been discussed in \cite{Seiberg:2023cdc}, it is possible to implement the $Z_{N}$ KW duality by shifting the parafermionic parity of $Z_{N}$ invariant sectors or shuffling the order and disorder operators. This gives concrete and visible implications of the KW duality (However it should be distinguished from the duality in a system with open boundaries where the duality can be implemented by the shift of topological defects or domain walls from one boundary to the other \cite{Frohlich:2004ef,Frohlich:2006ch,Schweigert:2007wd,Cobanera:2009as,Cobanera:2011wn,Cobanera:2012dc,Fukusumi:2020irh,Li:2023knf,Li:2023mmw}). This understanding may give a clue to establish a concrete understanding of simple current extended CFT based on the operator-state correspondence.

At this stage, one can introduce SD and define the correlation functions uniquely. Hence, under the bulk-edge correspondence, we conjecture that this procedure gives a canonical way to define the wavefunction of the topologically ordered system (hopefully even in general dimensions)\cite{Schoutens:2015uia}. We summarize the set of operators after the doubling trick as follows,
\begin{align}
\phi_{i,p}, \\
\{ \Phi_{a, p}\}_{SD}&=\sum_{p'}\frac{\phi_{a,p'+p}\phi_{\overline{a},-p'}} {\sqrt{N}}.
\end{align}
As one can see the $Z_{N}$ invariant sector and $Z_{N}$ noninvariant sector take different forms, and this is a remarkable consequence of $Z_{N}$ simple current extension. More mathematically, one can understand the $Z_{N}$ invariant sector is a sector refusing the $Z_{N}$ grading and this is a reason why the above two-point objects appear. Related discussion with emphasis on the entanglement struture in the Moore-Read states can be seen in \cite{Stern_2004}.
The same method may be applicable to symmetry-protected topological orders \cite{Scaffidi_2016} and this may be an interesting future problem.

\subsection{Comment on the difficulty to define chiral CFT and the corresponding topological order only from vertex operator algebra and singular vector}
 
In the previous sections, we have emphasized the SD of bulk CFT can give a consistent way to define a topological order and the corresponding wavefunctions. It may be worth noting that a chiral CFT in the existingliterature defined by vertex operator algebra (VOA) and the corresponding singular vector contains too many forms of correlation functions. In other words, one has to choose ``physical" correlation functions from such sets of correlation functions, but this ambiguity has rarely been noticed in the existing literature.

As a simplest example, let us consider the four-point function $\langle \psi(0)\psi (z) \psi (1) \psi (\infty) \rangle$ in the Ising or Majorana CFT. When one represents $\psi$ as $\phi_{(2,1)}$ by using the notation of the minimal $M(3,4)$ CFT, this results in a second-order differential equation, known as the BPZ equation. Hence there exist two solutions. However, one can represents $\psi$ as $\phi_{(1,3)}$ without sufficient information. This results in the third-order differential equation and there exist three solutions. This type of ambiguity in defining chiral correlation function has been introduced in the study of two-dimensional statistical models, typically percolation\cite{Watts:1996yh}, but this has never captured sufficient attention of condensed matter physicists\cite{Gori:2017cyq,Gori:2018gqx}.

However, we know the correlation function corresponding to the wavefunction of a topologically ordered system, Moore-Read state, is Pfaffian and the four-point function should be defined uniquely. Hence as a criterion to define the correlation function and the corresponding topological order uniquely, we propose CCFT/DCFT correspondence as a generalization of the CFT/TQFT in \cite{Fuchs:2002cm,Fuchs:2004dz,Vanhove:2018wlb} (as far as we know,  related subtelities to define topological order or TQFT have been pointed out only in \cite{Hung:2019bnq}).(After the submission of this work to arXiv, two works to construct group extension of chiral CFT has appeared\cite{Galindo:2024qzg,Gannon:2024tcl}. Their methods are based on VOA which are complementary to our method based on the fusion rings and CFT/TQFT.\footnote{ We thank authors of \cite{Galindo:2024qzg} for the helpful communications.})

\section{conclusion}
\label{conclusion}
In this work, the structure and interpretations of conformal blocks in the Ising CFT in the existing literature are clarified by using the rediscovered fermionized representation. Whereas the bulk correlation function can be constructed by using the operator-state correspondence (and the bootstrap technique), its chiral counterpart needs further condition CCFT/DCFT correspondence. The CCFT/DCFT correspondence can be considered as an operator version of bulk-edge correspondence by interpreting the CCFT as wavefunctions of topologically ordered systems. The appearance of Schottky double was emphasized in the author's previous work\cite{Fukusumi:2022xxe}, and related correspondence between bulk topological order and CCFT appeared ubiquitously in the existing literature\cite{Li_2008,Qi_2012,Cardy:2015xaa,Lencses:2018paa}.

We list a few open problems. First, it may be interesting to establish the concrete relationship between our method and the existing construction of the fusion algebra by orbifolding \cite{Dijkgraaf:1989hb}. In this work, we have mainly concentrated on the vacuum expectation value of the correlation functions which are insensitive to whether the representation is the bosonic or fermionic. However, when considering the torus CFT which inevitably contains topological defects or corresponding excited states, the results can depend on the representations as in \cite{Petkova:2009pe}.

Secondly, it seems interesting to consider the interpretations and realization of the second quantized expression of non-abelian anyon as we have discussed in Sec. \ref{Numerical_Moore-Read}. As can be seen easily, existing formalism like VOA and the corresponding calculations contain subtlety when considering its realization in a lattice model. Hence more direct definition of non-abelian anyon seems necessary and our method may give a clue in formulating them more evidently.

\begin{figure}[htbp]
\begin{center}
\includegraphics[width=0.5\textwidth]{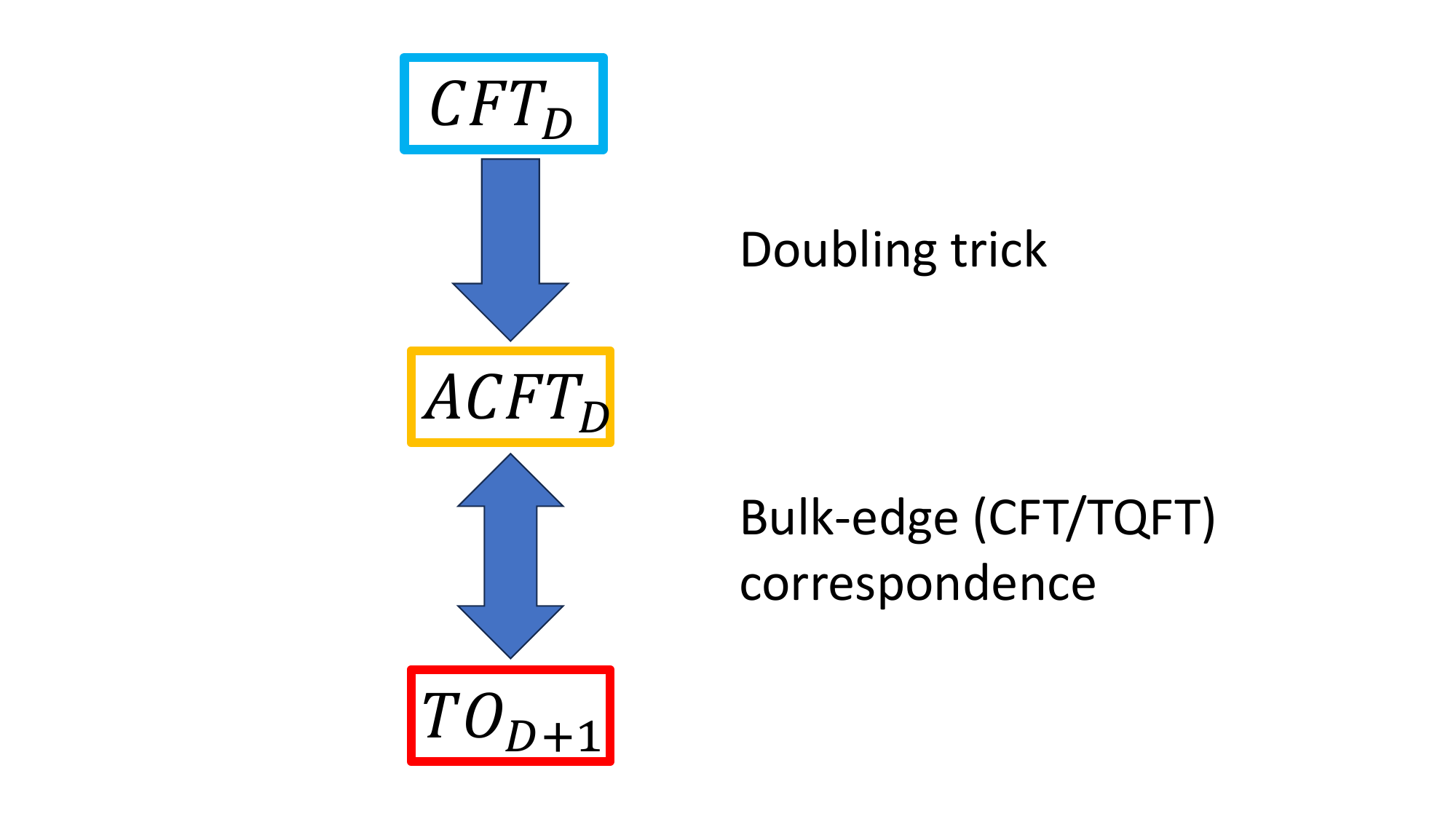}
\caption{Relations between $D$ dimensional bulk CFT, ACFT, and $D+1$ dimensional topological order (TO). For the $2+1$ dimensional topological order we have mainly discussed, the ACFT is called CCFT. It should be stressed that ACFT itself can be ``unphysical" to some extent depending on its anomaly, but can be interpreted as a ``physical" object in its correspondence to the TO. This gives a modern understanding of bulk-edge or CFT/TQFT correspondence, at the level of the correlation functions and the wavefunctions.}
\label{ACFT}
\end{center}
\end{figure}

Finally, as one may have already noticed, our discussion of CCFT/DCFT correspondence can be generalizable to the system with higher space-time dimensions by identifying them with BCFT.This implies the importance of studying BCFT, especially when considering the construction of wavefunction of topologically ordered systems in general space-time dimensions. As we have mentioned in Sec. \ref{exact_result}, we conjecture the following correspondence between correlation functions of bulk CFT and ACFT\cite{Nishioka:2022ook} and wavefunctions of topologically ordered systems (FIG.\ref{ACFT}).

 In this direction of research, it may be interesting to consider the relation between our method and the existing matrix product state and tensor network formulation of topologically ordered systems\cite{Zaletel_2012,Wu_2015,Wu_2014,Chen:2022wvy,Zeng:2023dzh} (We summarized related theoretical backgrounds in Appendix \ref{mathematical_literature}). For this purpose, the boundary conformal bootstrap technique can be fundamentally important\cite{Cardy:1991tv,Lewellen:1991tb,Herzog:2017xha,Bissi:2018mcq,Dey:2020jlc,Kusuki:2021gpt}.

\section{acknowledgement}
We thank Greg Henderson, Kansei Inamura, Kohki Kawabata, Yuya Kusuki, Hosho Katsura, Simon Lentner, Yutaka Matsuo, Steven Simon, Yuji Tachikawa, and Yunqin Zheng for the helpful comments and discussions. We especially thank Hosho Katsura for sharing his knowledge of literature and notifying us of the reference \cite{Ardonne:2010hj}. We also thank Ken Kikuchi for helpful comments on our draft and for notifying us of related ambiguities of fermionization. YF thanks many interesting lectures and  discussions in the three independent conferences, ``Symmetry 2024" at the United Kingdom, ``Conference on Recent Developments in Topological Quantum Field Theory" at China and ``Focus week on Non-equiribrium physics" at Japan. In particular, he thanks Jurgen Fuchs, Yuma Furuta, Yasuyuki Kawahigashi, Ingo Runkel, Christoph Schweigert, Kareljan Schoutens, Xiao-Gang Wen, and Masahito Yamazaki for helpful discussions on the aspects of category theory.
This work is supported by the National Research Foundation, Singapore under the NRF fellowship award (NRF-NRFF12-2020-005).

\appendix

\section{Descendant field, conformal tower and singular vector}
\label{CFT_structure}
In the main text, we have concentrated our attention on the correlation functions (and the corresponding wavefunctions of FQHE) constructed from the primary fields. Here, we comment on the correlation functions containing the descendant fields which can be constructed from those only with primary fields. By applying the identity coming from the mode expansion of a primary field labeled by $\alpha$, we can obtain those of the descendant fields and the corresponding wavefunctions.
\begin{equation}
L_{m}\phi_{\alpha}(\omega)\sim h_{\alpha} (m+1)\omega^{m} \phi_{\alpha} (\omega)+z^{m+1}\partial \phi (\omega)
\end{equation}
where $h_{\alpha}$ is the conformal dimension of the field $\phi_{\alpha}$ and $L_{m}$ is the generator of Virasoro algebra, $[L_{m},L_{n}]=(m-n)L_{m+N}+(c/12)\delta_{m+n,0}(m^{3}-m)$.

The following Ward-Takahashi identity may also be useful, 

\begin{equation}
\langle L_{m}\phi_{\alpha} \prod_{i}\phi_{\beta_{i}}(z_{i})\rangle =\mathcal{L}_{m}\langle \phi_{\alpha} \prod_{i}\phi_{\beta_{i}}(z_{i})\rangle,
\end{equation}
where $\mathcal{L}_{m}$ is,

\begin{equation}
\mathcal{L}_{m} = -\sum_{i}(m+1)h_{\beta_{i}}(z_{i}-\omega)^{m}+(z_{i}-\omega)^{m+1}\partial_{z_{i}}
\end{equation}

By applying the above operations to the CFT and $U(1)$ parts, one can obtain the wavefunctions and operators corresponding to the CFT characters. More detailed discussions can be seen in \cite{Milovanovic:1996nj}, for example.

Here we also note the explicit form of the BPZ equation of minimal conformal field theory (As a concise lecture note, see \cite{Ribault:2016sla}). The minimal conformal field theory $M(p,q)$ can be labeled by the coprime integer $p$, $q$. The central charge of the model is
\begin{equation}
c=1+6\left(b+b^{-1}\right)^{2}, \ b=i\frac{q}{p},
\end{equation}
where we have introduced the parameter $b$ following the notation in \cite{Ribault:2016sla}. In this model, there exist
degenerate fields $\phi_{(r,s)}$ labeled by two integer $(r,s)$ with conformal dimension $h_{(r,s)}=\left(\left(b+b^{-1}\right)^{2}-\left(rb+sb^{-1}\right)^{2}\right)/4$.This theory has remarkable singular vectors $\phi_{(2,1)}$ and $\phi_{(1,2)}$ with the following condition,
\begin{align}
\left( b^{2}L_{-1}^{2}+L_{-2}\right)|\phi_{(2,1)}\rangle&\in \text{null states}, \\
\left( b^{-2}L_{-1}^{2}+L_{-2}\right)|\phi_{(1,2)}\rangle&\in \text{null states},
\end{align}
{where the right hand side means the orthogonality of other states in the theory (conventionally, one can think these states are vanishing).
One can check the above equations by applying $L_{1}$ and $L_{2}$ to the lefthand side of them.
Hence, by replacing $L_{m}$ to $\mathcal{L}_{m}$ of the above equations by using the Ward-Takahashi Identity, one can obtain the second-order differential equations for the multipoint correlation functions containing  $\phi_{(2,1)}$ and $\phi_{(1,2)}$. These equations are the so-called BPZ equation and result in two conformal blocks that we have mainly discussed in the main text, by identifying the model as $M(3,4)$ and the conformal dimensions of the operators as $h_{(1,2)}=1/16$ and $h_{(2,1)}=1/2$.

The application of the BPZ equation to CCFT or BCFT appeared in the notable work  \cite{Cardy:1991cm} by Cardy and in the studies of related stochastic models, so-called Schramm-Loewner evolution, and can be seen in wide literature studying boundary critical phenomena (see review \cite{Bauer:2006gr} and reference therein, for example). Also, its application for the fractional quantum Hall system can be seen in \cite{Estienne:2010as}. We also note a recent application for the BPZ equation to a probabilistic construction of Liouville field theories \cite{ang2023liouville}.

\section{ U(1) charge conjugation}

We mainly considered the "electron" basis, with the quasihole $\{ e^{i\frac{r\varphi(\omega)}{\sqrt{q}}}\}$ and electron  $\{ e^{i\sqrt{q}\varphi(z)}\}$. However, one can consider the $U(1)$ charge conjugate basis,  $\{ e^{-i\frac{r\varphi(\omega)}{\sqrt{q}}}\}$ and $\{ e^{-i\sqrt{q}\varphi(z)}\}$. Applying this operation to the wavefunction and the corresponding wavefunction, nothing has changed. This is analogous to a kind of particle-hole symmetry, but this should be distinguished from it because particle-hole conjugate can change the structure of wavefunctions fundamentally. In the level of CFT, one can consider the correlation functions containing both  $\{ e^{\pm i\frac{r\varphi(\omega)}{\sqrt{q}}}\}$, but this seems to be difficult to realize in the microscopic models because of the singularities caused by condensation of the charge positive and negative particles. Instead, the realization of such wavefunction with both quasielectron and quasihole has been proposed in the ``lattice" FQHE systems\cite{Manna:2018lqg,Glasser:2015fla}. It should be noted that, in this language, antiholomorphic variables $\overline{\omega}$ and $\overline{z}$ do not appear.

Also related to the appearance of the antiholomorphic part of wavefunction, this seems to correspond to the Schottky double in BCFT, because the antiholomorphic fields seem to be related to the holomorphic fields. (In full CFT, $\langle \overline{\phi}_{\alpha} (\overline{\omega})\prod_{i}\phi_{\beta_{i}}(z_{i}) \rangle$ is $0$. However, this can become nonzero by applying Schottky double, because some identification, like $\overline{\phi}_{\alpha}=\phi_{\alpha'}$, appears.)

\section{Set of correlation functions}

In this section, we note a simple set of correlation functions labeled by edge states or operators at $\infty$. By multiplying $U(1)$ part, these correlation functions produce a wavefunction of (generalized) Moore-Read state which we have numerically tested in the main text.
\subsection{ Identity operator at the edge}
The correlation function only with $\psi$ field is,
\begin{equation}
\langle \prod_{i}^{N}\psi (z_{i})\rangle=\text{Pf}\left[ \frac{1}{(z_{i}-z_{j})}\right]_{i,j}
\end{equation}
where we have taken $N$ as an even integer.
The simplest correlation function which contains $\sigma$ field (or SD of $\sigma_{\text{Bulk}}$ or $\mu_{\text{Bulk}}$) is,
\begin{equation}
\begin{split}
&\langle \sigma_{\text{Bulk}} (\omega_{1},\omega_{2}) \prod_{i}^{N}\psi (z_{i})\rangle_{\text{SD}} \\
&=\text{Pf}\left[ \frac{1}{(z_{i}-z_{j})} \left(\sqrt{\frac{z_{i}-\omega_{1}}{z_{j}-\omega_{1}}}\sqrt{\frac{z_{j}-\omega_{2}}{z_{i}-\omega_{2}}}+\sqrt{\frac{z_{j}-\omega_{1}}{z_{i}-\omega_{1}}}\sqrt{\frac{z_{i}-\omega_{2}}{z_{j}-\omega_{2}}}\right)\right]_{i,j} \\
&\times (\omega_{1}-\omega_{2})^{-1/8}
\end{split}
\end{equation}
where we have taken $N$ as an even integer. By taking limit and applying fusion rules, one can construct a class of correlation functions systematically.

By fusing one $\psi$ to $\sigma$, one can obtain the electron odd wavefunction,
\begin{equation}
\begin{split}
&\langle \mu_{\text{Bulk}} (\omega_{1},\omega_{2}) \prod_{i}^{N}\psi (z_{i})\rangle_{\text{SD}} \\
&=\lim_{z_{N+1}\rightarrow \omega_{1}} \sqrt{z_{N+1}-\omega_{1}} \\
&\text{Pf}\left[ \frac{1}{(z_{i}-z_{j})} \left(\sqrt{\frac{z_{i}-\omega_{1}}{z_{j}-\omega_{1}}}\sqrt{\frac{z_{j}-\omega_{2}}{z_{i}-\omega_{2}}}+\sqrt{\frac{z_{j}-\omega_{1}}{z_{i}-\omega_{1}}}\sqrt{\frac{z_{i}-\omega_{2}}{z_{j}-\omega_{2}}}\right)\right]_{i,j} \\
&\times (\omega_{1}-\omega_{2})^{-1/8} \\
&(= \lim_{z_{N+1}\rightarrow \omega_{1}}\sqrt{(z_{N+1}-\omega_{1})}\langle \sigma_{\text{Bulk}} (\omega_{1},\omega_{2}) \prod_{i}^{N+1}\psi (z_{i})\rangle_{\text{SD}})
\end{split}
\end{equation}
where we have taken $N$ as an odd integer, and introduced $z_{N+1}$ to define Pfaffian.
The above gives a basic structure of correlation functions.

\subsection{Majorana Fermion at the edge}
The correlation function only with $\psi$ field is,
\begin{equation}
\langle \psi(\infty) \prod_{i}^{N}\psi (z_{i})\rangle=\lim_{z_{N+1}\rightarrow \infty}z_{N+1}\text{Pf}\left[ \frac{1}{(z_{i}-z_{j})}\right]_{i,j}
\end{equation}
where we have taken $N$ as an odd integer. The factor $z_{N+1}$ before the Pfaffian part comes from the normalization of states at $\infty$.
Similar to the $I(\infty)$ case, one can obtain the following wavefunctions,
\begin{equation}
\begin{split}
&\langle \psi (\infty) \sigma_{\text{Bulk}} (\omega_{1},\omega_{2}) \prod_{i}^{N}\psi (z_{i})\rangle_{\text{SD}} \\
&=\lim_{z_{N+1} \rightarrow \infty} z_{N+1} \\
&\text{Pf}\left[ \frac{1}{(z_{i}-z_{j})} \left(\sqrt{\frac{z_{i}-\omega_{1}}{z_{j}-\omega_{1}}}\sqrt{\frac{z_{j}-\omega_{2}}{z_{i}-\omega_{2}}}+\sqrt{\frac{z_{j}-\omega_{1}}{z_{i}-\omega_{1}}}\sqrt{\frac{z_{i}-\omega_{2}}{z_{j}-\omega_{2}}}\right)\right]_{i,j} \\
&\times (\omega_{1}-\omega_{2})^{-1/8}
\end{split}
\end{equation}
where we have taken $N$ as an odd integer.
From a similar argument, for $N$ even, one can obtain,
\begin{equation}
\begin{split}
&\langle \psi(\infty) \mu_{\text{Bulk}} (\omega_{1},\omega_{2}) \prod_{i}^{N}\psi (z_{i})\rangle_{\text{SD}} \\
&=\lim_{z_{N+1}\rightarrow \omega_{1}, z_{N+2}\rightarrow \infty} z_{N+2} \sqrt{z_{N+1}-\omega_{1}} \\
&\text{Pf}\left[ \frac{1}{(z_{i}-z_{j})} \left(\sqrt{\frac{z_{i}-\omega_{1}}{z_{j}-\omega_{1}}}\sqrt{\frac{z_{j}-\omega_{2}}{z_{i}-\omega_{2}}}+\sqrt{\frac{z_{j}-\omega_{1}}{z_{i}-\omega_{1}}}\sqrt{\frac{z_{i}-\omega_{2}}{z_{j}-\omega_{2}}}\right)\right]_{i,j} \\
&\times (\omega_{1}-\omega_{2})^{-1/8} 
\end{split}
\end{equation}

\subsection{Chiral order operator at the edge}
One can see the correlation function can be defined by taking limit $\omega_{2}\rightarrow \infty$ for $\langle \sigma_{\text{Bulk}} (\omega_{1},\omega_{2}) \prod_{i}^{N}\psi (z_{i})\rangle_{\text{SD}} $ or $\langle \mu_{\text{Bulk}} (\omega_{1},\omega_{2}) \prod_{i}^{N}\psi (z_{i})\rangle_{\text{SD}} $.

\begin{equation}
\begin{split}
&\langle \sigma_{\text{Bulk}} (\omega_{1},\infty) \prod_{i}^{N}\psi (z_{i})\rangle_{\text{SD}} \\
&=\text{Pf}\left[ \frac{1}{(z_{i}-z_{j})} \left(\sqrt{\frac{z_{i}-\omega_{1}}{z_{j}-\omega_{1}}}+\sqrt{\frac{z_{j}-\omega_{1}}{z_{i}-\omega_{1}}}\right)\right]_{i,j} 
\end{split}
\end{equation}
for $N$ even, and
\begin{equation}
\begin{split}
&\langle \mu_{\text{Bulk}} (\omega_{1},\infty) \prod_{i}^{N}\psi (z_{i})\rangle_{\text{SD}} \\
&=\lim_{z_{N+1}\rightarrow \omega_{1}} \sqrt{z_{N+1}-\omega_{1}}\\
&\times \text{Pf}\left[ \frac{1}{(z_{i}-z_{j})} \left(\sqrt{\frac{z_{i}-\omega_{1}}{z_{j}-\omega_{1}}}+\sqrt{\frac{z_{j}-\omega_{1}}{z_{i}-\omega_{1}}}\right)\right]_{i,j} 
\end{split}
\end{equation}
for $N$ odd.

The general form by using Haffian can be seen in the discussion around Eq. (20) in \cite{Ardonne:2010hj}

\section{Reconstructing a bulk theory from the composition of the chiral and antichiral theories}
\label{reconstruction}

In this section, we demonstrate algebraic details of the construction of bulk operators in Majorana CFT from the combination of the chiral and antichiral CFTs. The arguments are totally algebraic, so readers unfamiliar with recent sophisticated technical (or mathematical) arguments can follow. This is one of the benefits of our approach.  

First of all, we assume the following set of fusion rules (or algebraic data) as in the main text,
\begin{align}
\psi\times \psi&= I, \quad e\times m=\psi, \\
\psi\times e&= m, \quad\psi\times m= e, \\
e \times e&= m\times m =I.
\end{align}
We also assume the same fusion rule for the antichiral fields. Then we introduce the following representation of the bulk operators,
\begin{align}
\epsilon&= \psi \overline{\psi}, \\
\sigma_{\text{Bulk}}&= \frac{e\overline{e}+m\overline{m}}{\sqrt{2}}, \\
\mu_{\text{Bulk}}&= \frac{e\overline{m}+m\overline{e}}{\sqrt{2}}, \\
\end{align}

By assuming the chiral and antichiral decomposition of the bulk operators $\{ I,\psi, \overline{\psi}, \epsilon, \sigma_{\text{Bulk}}, \mu_{\text{Bulk}} \}$, one can demonstrate the above bulk operators satisfying the fusion rule of bulk CFT. For example, one can show the following equation corresponding to the parity shift operation in \cite{Runkel:2020zgg}
\begin{equation}
\psi \times \sigma_{\text{Bulk}} =\psi \times \frac{e\overline{e}+m\overline{m}}{\sqrt{2}}=\frac{e\overline{m}+m\overline{e}}{\sqrt{2}}=\mu_{\text{Bulk}}
\end{equation}
and the nonabelian fusion rule,
\begin{equation}
 \sigma_{\text{Bulk}} \times \sigma_{\text{Bulk}} =\frac{e\overline{e}+m\overline{m}}{\sqrt{2}} \times \frac{e\overline{e}+m\overline{m}}{\sqrt{2}}=I+\epsilon .
\end{equation}

By studying the $Z_{2}$ structure of $I, \psi, e,m$ and their antiholomorphic counterparts, one can obtain the fusion rule of the bulk operators of the Majorana CFT in the main text. It may be worth stressing that the unconventional factor $1/\sqrt{2}$ is unavoidable for this chiral and antichiral decomposition. To some extent, this is the reason why recent literature on the fields lacks the chiral and antichiral decomposition of the fields in the Ramond sector (whereas this has not captured sufficient attention in the communities regardless of the appearance in the notable review \cite{Ginsparg:1988ui}).

We also note another choice of the basis in \cite{Ginsparg:1988ui} coming from the reduction of the four-dimensional space of the R sector to two-dimensional one. This representation can be realized by changing the definition of the bulk fields in the R sector,
\begin{align}
\epsilon&= \psi \overline{\psi}, \\
\sigma'_{\text{Bulk}}&= \frac{e\overline{e}-m\overline{m}}{\sqrt{2}}, \\
\mu'_{\text{Bulk}}&= \frac{e\overline{m}-m\overline{e}}{\sqrt{2}}.
\end{align}
However, when considering the fusion rule of the bulk operator, the bulk theory cannot be the same. Related discussion and the interpretation of this algebraic structure have been summarized in other works by the first author\cite{Fukusumi:2024cnl}.

It should be noted that the bosonic bulk CFT, $\{ I,\epsilon, \sigma_{\text{Bulk}}\}$, cannot be the same as the (naive) product of bosonic chiral and antichiral CFTs, at the level of the operator. It is necessary to consider some condensation in the theory, and this is the reason why we have introduced the notation $\sigma_{\text{Bulk}}$, to distinguish it from $\sigma \overline{\sigma}$. By using the terminology in category theory, this condensation $\sigma \overline{\sigma}$ to   $\sigma_{\text{Bulk}}$ is called \emph{abstract nonsense}\cite{Kong:2013aya,etingof2009weaklygrouptheoreticalsolvablefusion}. In other words, by considering the $Z_{2}$ extension, one can specify the chiral and antichiral decomposition of the bulk CFT more intuitively. This is natural because the classification of CFTs has been achieved based on finite group symmetries. We note a recent review\cite{Cappelli:2009xj} and work \cite{Gannon:2023udg}, for the convenience of the readers.

\section{Algebraic structure and asymptotic behavior of chiral conformal field theory under simple current extension}
\label{algebra_OPE}
Here we propose the structure of chiral conformal field theory after simple current extension and their implication for the assymptotic behavior under OPE. The algebraic structure can be fundamental for the MPS construction of the model and the assymptotic behavior is fundamental to consider statistics of quasiholes or anyons.

First, let us assume the following asymptotic expansion for the fields $\phi$ labeled by $\alpha, \beta, \gamma$,
\begin{equation}
\phi_{\alpha}(z) \phi_{\beta}(z')\sim \sum_{\gamma} C^{\gamma}_{\alpha \beta} (z-z')^{h_{\gamma}-h_{\alpha}-h_{\beta}}\phi_{\gamma}(z')
\end{equation}
where $h$ is the conformal dimension of the field and $C$ is the OPE coefficient which is consistent with the fusion rule. Even after simple current expansion, this can be well defined, but it is necessary to introduce two-point objects for consistency with existing formalism.

In the main text, we have assumed the asymptotic expansion and compared the result with the existing forms which can be derived from the BPZ equation.

For the readers unfamiliar with this kind of asymptotic analysis, we note the relevant forms constructed from one-point objects,
\begin{align}
\psi(z)\psi(z')&\sim(z-z')^{-1}, \\
e(z)e(z')&\sim(z-z')^{-\frac{1}{8}}, \\
m(z)m(z')&\sim(z-z')^{-\frac{1}{8}}, \\
e(z)m(z')&\sim(z-z')^{\frac{3}{8}}\psi(z'), \\
\psi(z)e(z')&\sim(z-z')^{-\frac{1}{2}}m(z'), \\
\psi(z)m(z')&\sim(z-z')^{-\frac{1}{2}}e(z'), \\
\end{align}

For the two-point objects, one can obtain the following asymptotic expansion,
\begin{align}
\{\sigma_{\text{Bulk}}(z,z')\}_{\text{SD}}&\sim \sqrt{2} (z-z')^{-\frac{1}{8}}, \\
\{\mu_{\text{Bulk}}(z,z')\}_{\text{SD}}&\sim \sqrt{2} (z-z')^{\frac{3}{8}}\psi (z'),
\end{align}
Hence the two objects result in two different asymptotic behaviors. These two behaviors come from the doubling trick we have applied. (One can exchange the role of order and disorder operators, but their asymptotic behaviors are different because of the doubling trick. This is a reason why analysis of the BCFT becomes nontrivial compared with the bulk CFT.)

For simplicity, we note the analysis of the following four-point operators,
\begin{align} 
&\{\sigma_{\text{Bulk}}(z_{1},z_{2})\sigma_{\text{Bulk}}(z_{3},z_{4})\}_{\text{SD}} \\
&\{\mu_{\text{Bulk}}(z_{1},z_{2})\mu_{\text{Bulk}}(z_{3},z_{4})\}_{\text{SD}}
\end{align}
For these operators, one can consider the following two asymptotics,
\begin{align}
z_{1}&\sim z_{2}, z_{3}\sim z_{4}, \\
z_{1}&\sim z_{3}, z_{2}\sim z_{4}.
\end{align}
The former gives the following asymptotic behavior,
\begin{align} 
\sigma_{\text{Bulk}}(z_{1},z_{2})\sigma_{\text{Bulk}}(z_{3},z_{4}) &\sim 2 (z_{1}-z_{2})^{-\frac{1}{8}} (z_{3}-z_{4})^{-\frac{1}{8}}, \\
\mu_{\text{Bulk}}(z_{1},z_{2})\mu_{\text{Bulk}}(z_{3},z_{4}) &\sim 2 (z_{1}-z_{2})^{\frac{3}{8}} (z_{3}-z_{4})^{\frac{3}{8}}\psi(z_{2})\psi(z_{4}).
\end{align}
From these behaviors, one can detect the conformal block easily from the given forms of correlation functions in \cite{Ardonne:2010hj}. The latter asymptotics are more complicated. Combinations of the exponents $(z_{1}-z_{3})^{\frac{3}{8}}$, $(z_{1}-z_{3})^{-\frac{1}{8}}$, $(z_{2}-z_{4})^{\frac{3}{8}}$, and $(z_{2}-z_{4})^{-\frac{1}{8}}$ appears, and this is consistent with the exiting conformal blocks.

\section{Mathematical frameworks (still under development)}
\label{mathematical_literature}
In this section, we summarize the recent development of mathematical frameworks to formulate CCFT corresponding to FQHE. Related earlier research directions for CCFT and their relation to the conformal net have been summarized in the works by Kawahigashi and his collaborator\cite{Kawahigashi:2015zha,Kawahigashi:2021hds,Evans:2023nbp}. Before going into the details, we stress that the existing well-known framework, modular tensor category (MTC), does not correspond to an FQHE in general. The resultant theory can be outside of existing categorical frameworks because of the appearance of the two-point objects in the main text. In other words, our formalism is an extension of the construction of chiral CFTs by using CFT/TQFT or quantum 6j-symbol \cite{Reshetikhin:1990pr,Turaev:1992hq,Fuchs:2002cm,Fuchs:2004dz}. Recently, these studies have captured the attention of condensed matter theorists because of their close connection to Tensor-Network \cite{Vanhove:2018wlb,Lootens:2020mso,Chen:2022wvy,Eck:2024myo,Ueda:2024jzj}.

First, we remark on the exceptional well-organized cases, in the studies of vertex operator algebra and its simple current extension in close connection to Monstrous-Moonshine \cite{Huang:1994sx,yamauchi2003modulecategoriessimplecurrent,Lam2004VertexOA}\footnote{We thank Yuji Tachikawa for pointing out this research direction}. The $E_{8}$ models are typical examples and we note recent works \cite{BoyleSmith:2023xkd,Rayhaun:2023pgc,Hohn:2023auw} in this research direction. Application of such specialized theories can be seen in the recent literature on the bosonic FQHE\cite{Lim:2022vee}. More recently, related research for the ``anomalous" model has appeared\cite{Galindo:2024qzg,bischoff2021computingfusionrulesspherical}. 

Second, we note recent studies on the $G$-extension of category theory where $G$ is a group acting on a category theory. The simple current extension of a CFT in the main text is a typical example and the significance of this categorical framework in condensed matter has been revisited in \cite{Barkeshli:2014cna,Lan_2016,Lan2016ModularEO,Cho:2022kzf,Seo:2023gnh} for example. The framework $G$-graded category in \cite{etingof2009weaklygrouptheoreticalsolvablefusion} is the most general at this stage, but we note that a naive application of $Z_{N}$-extension of $Z_{N}$ symmetric modular tensor category (or bosonic CCFT) does not correspond to the models we have studied in the main text. To obtain a categorical analog of the $Z_{N}$ extended chiral CFT in this work, it is necessary to obtain $Z_{N}$ extension of full CFTs (or $Z_{N}$ extension of spherical fusion category \cite{bischoff2021computingfusionrulesspherical,davydov2021braidedpicardgroupsgraded}) and consider their condensation to the chiral CFT \cite{Huang:2023pyk,Fukusumi:2024cnl}. However, it should be remarked that this is just an algebraic or generalized symmetry based expression of the bulk-edge correspondence in existing literature. The appearance of CCFT (or BCFT) from the bulk CFT can be seen in \cite{Qi_2012,Cardy:2017ufe}. As the authors have revisited in \cite{Fukusumi:2022xxe}, the corresponding phenomena have already been known in the studies of massive integrable field theories and the corresponding lattice models in the later 1980s\cite{Date:1987zz,Saleur:1988zx,Foda:2017vog}. We also remark that related proposals \cite{Huang:2023pyk,Wen:2024udn,Bhardwaj:2024ydc,Huang:2024ror} to obtain the structure of the chiral CFT appeared before and after the submission of this work (However, whereas these works are based on a general formalism,  ``topological holography" or ``sandwich construction" \cite{Moradi:2022lqp,Bhardwaj:2023bbf}, they contain misleading expressions at this stage. For example, they fail to distinguish order and disorder fields (and their chiral analogs) and use notations resulting in $1\times 1 \neq 1$. We also note that the terminology ``topological holography" has appeared earlier in a different context\cite{Husain:1998vz}.). As more concrete works for the simple current extension of CCFT, we cite the established review \cite{Ginsparg:1988ui} for the $Z_{2}$ case and the recent work \cite{Fukusumi:2024cnl} by the first author for more general cases again.

Finally, we note literature on code CFTs \cite{Almheiri:2014lwa,Pastawski:2015qua} which has a connection to information science and their application to simple current extension\cite{Dymarsky:2020qom,Furuta:2022ykh,Kawabata:2022jxt,Furuta:2023xwl,Kawabata:2023nlt,Ando:2024gcf,Kawabata:2024gek}. The interpretation of the two-point objects in the main text in this context is an interesting open problem. We expect this may reveal fundamental information-theoretic aspects of TOs with a close connection to their properties of entanglement\footnote{We comment on the analogy between the quantum stabilizer code and the nontrivial GSO projection in the main text which reduces four-dimensional Ramond sectors to two-dimensional ones. We thank Kohki Kawabata for pointing out this. We also comment on the related earlier study in the condensed matter theory community in \cite{Stern_2004}. We thank Steven Simon for notifying the literature \cite{Stern_2004}.}.

\appendix

\bibliographystyle{ytphys}
\bibliography{Cardy_v3}

\end{document}